\documentclass[aps,prb,twocolumn,superscriptaddress,showpacs,preprintnumbers]{revtex4-1}

\usepackage{graphicx}

\begin{document}


\preprint{Published in Phys. Rev. B 87, 035123 (2013) \copyright American Physical Society}

\title{Unoccupied electronic band structure of the semi-metallic Bi(111) surface probed with two-photon photoemission}


\author{Christopher Bronner}
    \email{bronner@zedat.fu-berlin.de}
\affiliation{Freie Universit\"at Berlin, Fachbereich Physik, Arnimallee 14, 14195 Berlin, Germany}

\author{Petra Tegeder}
\affiliation{Freie Universit\"at Berlin, Fachbereich Physik, Arnimallee 14, 14195 Berlin, Germany}
\affiliation{Ruprecht-Karls-Universit\"at Heidelberg, Physikalisch-Chemisches Institut, Im Neuenheimer Feld 253, 69120 Heidelberg, Germany}


\date{published in Phys. Rev. B on 17 January 2013}

\begin{abstract}
While many photoemission studies have dealt with both the bulk band structure and various surface states and resonances, the unoccupied electronic structure above the Fermi level of the Bi(111) surface has not yet been measured directly although understanding of this model semi-metal is of great interest for topological insulators, spintronics and related fields. We use angle-resolved two-photon photoemission to directly investigate the occupied and unoccupied $p$-bands of Bi, including the bulk hole pocket at the $T$-point, as well as the image potential states and surface states of Bi(111).
\end{abstract}

\pacs{71.20.Gj, 73.20.At}

\maketitle

\section{Introduction}

\begin{figure}[htbp]
\includegraphics{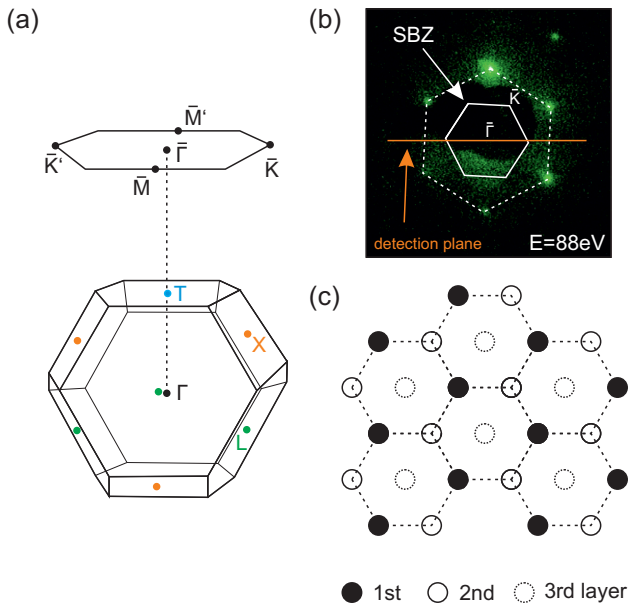}
\caption{(a) Brillouin zone and surface Brillouin zone (SBZ) of the rhombohedral Bi lattice together with important high-symmetry points. (b) LEED image obtained at an electron energy of 88~eV. The SBZ deduced from the LEED spots is shown in the image as well as the detection plane in our angle-resolved photoemission measurements. (c) Top-view of the Bi(111) surface. The hexagonal planes are stacked in an ABC fashion.\label{fig:Fig1-structure}}
\end{figure}

As a group-V element, Bi has an electronic structure which is dominated by its three valence electrons in the $6p^{3}$ orbitals. If Bi crystalized in a simple fcc crystal structure with five electrons per unit cell, like Pb or many transition metals, it would have an odd number of electrons per unit cell and thus it would be a metal due to a partially filled \emph{p}-band. Since Bi rather crystallizes in a rhombohedral structure (see Fig. \ref{fig:Fig1-structure}) which corresponds to two fcc sub-lattices that are slightly shifted with respect to each other along the [111] direction, the unit cell contains ten valence electrons which would allow Bi to be an insulator with three filled \emph{p}-bands and three empty \emph{p}-bands. However, Bi actually is a semi-metal, which is caused by the fact that, out of eight \emph{L}-points in an fcc Brillouin zone, in the rhombohedral structure the two on the [111] axis are inequivalent to the other six (and referred to as \emph{T}-points). This lattice distortion leads to electron pockets at the \emph{L}-points and hole pockets at the \emph{T}-points, which give rise to conducting carriers at the Fermi level and make Bi semi-metallic.\cite{{Hofmann2006},{Ast2002},{Ast2004}}
Besides the bulk bands, a rich variety of surfaces states and resonances have been found at Bi(111) and other low-index faces of Bi,\cite{{Jezequel1986},{Hengsberger2000},{Ast2001},{Koroteev2004},{Hofmann2006}} partly residing in the first bilayer.\cite{Ast2003} In fact, they form multiple electron and hole pockets and thereby account for the majority of carriers at the Fermi level. The surfaces of Bi therefore have a higher charge carrier density than the semi-metallic bulk. Furthermore, due to the loss of inversion symmetry at the surface and the importance of relativistic effects in this heavy element, spin-orbit interaction plays a crucial role for the surface states, which are spin-split by more than 100~meV.\cite{{Ohtsubo2012},{Koroteev2004}}

Particularly the spin-polarized surface states are very attractive for the use in spintronics, i.e. electronic devices which operate not only using the charge of the electrons but also their spin. For such applications, e.g. sources for spin-polarized electron currents and spin-filters have been proposed.\cite{{Datta1990},{Koga2002}} Unlike semiconductor hetero-structures, Bi surfaces are especially suitable in such devices not only because of the large energetic gap between the spin-split states (which can not be overcome thermally) but also because they contribute much more to the Fermi surface than the spin-degenerate bulk states.
Another interesting property of Bi is the emergence of superconductivity in rhombohedral clusters of only a few nanometers in size.\cite{{Vossloh1998},{Weitzel1991}} This behavior has been associated with the increased density of states at the Fermi level caused by the surfaces states and resonances and their low effective masses.\cite{{Patthey1994},{Hengsberger2000}}
The key role of Bi in several variants of topological insulators also fosters the broad interest in its electronic structure: Bi itself supports topologically metallic edge states on the (114)-face\cite{Wells2009} and it is the prime constituent of ${\rm{Bi}}_{1-x}{\rm{Sb}}_{x}$ (the structure of which is very similar to pristine Bi)\cite{Hsieh2008} and more complex layered systems such as ${\rm{Bi}}_{2}{\rm{Se}}_{3}$ and ${\rm{Bi}}_{2}{\rm{Te}}_{3}$ which are currently studied intensively.\cite{Zhang2009}
More generally, Bi and especially its (111)-surface can be considered an interesting model system for surfaces and interfaces because (i) the valence electronic structure is relatively simple, (ii) the surface states can be considered a quasi-two-dimensional metal and (iii) the surface exhibits neither a reconstruction nor dangling bonds since the (111)-surface lies in the natural cleavage plane.

Using angle-resolved photoelectron spectroscopy (ARPES) from the (111)-surface, the bulk valence electronic structure, i.e. the occupied \emph{p}-bands, has been studied in great detail in the past three decades. Three bands are observed which are each separated by a gap. Their dispersion has been investigated both along the $\Gamma T$-line\cite{{Jezequel1986},{Jezequel1997},{Ast2004}} which is parallel to the surface normal as well as parallel to the surface along the $\bar\Gamma\bar M$- and $\bar\Gamma\bar K$-lines of the surface Brillouin zone (SBZ).\cite{{Thomas1999},{Tanaka1999}} Along the surface, the \emph{p}-bands disperse with a negative effective mass  around the $\bar\Gamma$-point in an energy range from the Fermi level down to approximately 5~eV.\cite{{Thomas1999},{Tanaka1999}} A careful analysis of the dispersion perpendicular to the surface shows that the highest \emph{p}-band crosses the Fermi level at the \emph{T}-point and thus forms a hole pocket.\cite{Ast2004}
Within the energy range of the occupied \emph{p}-bands, at around $-0.4$~eV\cite{Ast2004} and $-0.6$~eV,\cite{Ast2003} two features that originate from the surface have been observed which disperse around the center of the SBZ with a negative effective mass\cite{{Ast2002},{Jezequel1986},{Patthey1994}} and threefold symmetry in the surface plane indicating surface resonance (rather than surface state) character.\cite{Ast2003}
In a narrow region around the Fermi level, several features associated with the surface are found which form electron pockets at the $\bar\Gamma$-point\cite{{Ast2001},{Ast2003},{Ohtsubo2012},{Koroteev2004}} and the $\bar M$-point\cite{{Hengsberger2000},{Ast2003}} as well as radial, droplet-shaped hole pockets.\cite{{Hengsberger2000},{Ast2001},{Ast2003},{Ohtsubo2012},{Koroteev2004}} All of these features belong to a spin-split surface state (which however has surface resonance character at the high symmetry points).\cite{Koroteev2004} The origin of this state in the first bilayer is demonstrated by its sixfold symmetry.\cite{{Ast2003},{Ast2003a}} Note that besides the electron and hole pockets induced by these surface bands, the (less pronounced) pockets of the bulk are still present. For example, at the $\bar\Gamma$-point, the electron pocket of the surface state lies concentrically with the hole pocket of the bulk which is found at the \emph{T}-point.\cite{Ast2001}

Unlike for the occupied states, studies on unoccupied electronic states are very rare. Using two-photon photoemission (2PPE), the image-potential state (IPS) of the Bi(111) surface lying 3.57~eV above the Fermi level has been studied in detail.\cite{Muntwiler2008}
Another 2PPE study has shown direct spectroscopic evidence of the spin-split surface state up to approximately 200~meV above the Fermi level.\cite{Ohtsubo2012} Using terahertz radiation, information could be gained on the dynamics of electrons excited into the conduction band but the method doesn't allow direct spectroscopic observation of the unoccupied bands and their absolute energetic position with respect to the Fermi level.\cite{Timrov2012}
A number of theoretical studies have dealt with both the bulk and the surface electronic structure of Bi using various methods such as pseudopotentials,\cite{Golin1968} first principles calculations\cite{Gonze1990} or tight-binding models.\cite{Liu1995} They all make predictions about the unoccupied band structure which, to the authors' knowledge, have not been specifically addressed experimentally so far.

In this paper, we briefly revisit the occupied electronic structure based on ultraviolet photoemission (UPS) measurements. We find a well-known surface resonance and observe two features originating from the occupied $p$-bands which might be influenced also by a second surface resonance. Using angle-resolved 2PPE we investigate the electronic structure of the unoccupied states. All three unoccupied $p$-bands are observed in the spectra as well as a final state above the vacuum level, two IPS and the bulk hole pocket.

\section{Experimental methods\label{sect:experimental}}

Two-photon photoemission (2PPE) has proven to be a powerful method to study both occupied and unoccupied electronic states at surfaces.\cite{{Weinelt2002},{Petek1997},{Zhu2004},{Gudde2006}} 2PPE is a surface-sensitive pump-probe technique using two ultrashort laser pulses with equal (one-color 2PPE, 1C-2PPE) or different (two-color 2PPE, 2C-2PPE) photon energies below the work function $\Phi$ of the sample. The first (pump) pulse excites an electron from an occupied electronic state to an unoccupied state. The probe pulse then emits the electron and its kinetic energy $E_{\rm{kin}}$ is measured.
Features of a 2PPE spectrum may arise from occupied states (probed with a two-photon process via a virtual intermediate state) or from unoccupied states. To identify the nature of a feature, the peak position in the spectrum can be monitored while varying the photon energy by $\Delta h\nu$.\cite{{Bronner2011},{Bronner2012}} If we consider a 1C-2PPE experiment, a peak originating from an unoccupied state will shift by $\Delta h\nu$. In contrast, a peak arising from an occupied state below the Fermi level $E_{\rm{F}}$ shifts by $2\Delta h\nu$ in energy. Note that this identification of states strictly is only applicable to states without dispersion perpendicular to the surface such as surface states or adsorbate-induced states. Otherwise the energy shift caused by the change in photon energy is not distinguishable from possible energy shifts which are due to the change of the probed state's binding energy because the probed point in reciprocal space varies with the photon energy.\cite{{Leyssner2010},{Hagen2010}} Because of the ambiguity of a peak's origin, 2PPE spectra are displayed versus the final state energy $E_{\rm{Final}}-E_{\rm{F}}=E_{\rm{kin}}+\Phi$, from which the binding energy can be conveniently obtained by subtracting the photon energy once or twice for unoccupied and occupied states, respectively.
In order to measure the dispersion of a state, angle-resolved (AR-2PPE) measurements can be performed by rotating the sample in front of the electron analyzer. The peak position is recorded as a function of momentum parallel to the surface,
\begin{equation}
k_{\parallel}=\sqrt{ \frac{2 m_{\rm{e}} E_{\rm{kin}}}{\hbar ^ 2}} \cdot \sin{\vartheta},
\end{equation}
which is altered with the emission angle $\vartheta$. The laser light is incident at the surface under an angle of 45$^\circ$ (in normal emission) and in the following, positive values for $k_{\parallel}$ refer to the situation, where the sample is rotated toward the beam.
If the probe pulse is delayed with respect to the pump pulse, the ultrafast dynamics of an unoccupied state can be studied. The delay is introduced by varying the path length of one beam. Depending on the sign of the pump-probe delay in a time-resolved 2C-2PPE experiment, either pulse can be pump or probe.

Femtosecond laser pulses at 800~nm are generated in a Ti:Sapphire oscillator and amplified in a regenerative amplifier at 300~kHz. This fundamental light can be frequency-doubled by second harmonic generation in BBO crystals once (400~nm) or twice (200~nm) or it can be converted to visible pulses using an optical parametric amplifier (OPA). These visible pulses can also be frequency-doubled to yield tunable ultraviolet light. Beams are p-polarized when incident on the sample.
The Bi(111) single crystal is mounted in an ultrahigh vacuum (UHV) chamber on a flow cryostat equipped with resistive heating. Besides a sputter gun, the chamber is equipped with a low-energy electron diffraction (LEED) apparatus and a time-of-flight (TOF) electron spectrometer.

We prepared the single crystal by routine cycles of Ar$^{+}$ sputtering (900~V) and annealing (410~K, 10~min). The preparation was checked by LEED where a sharp sixfold structure was observed and by comparison of the work function and image potential state with another 2PPE study.\cite{Muntwiler2008} Using LEED, we could determine the crystal orientation and verify that the sampling direction in the surface Brillouin zone is $\bar K'\bar\Gamma\bar K$ (see Fig. \ref{fig:Fig1-structure}).

Direct photoemission experiments, i.e. UPS, were conducted using the twice frequency-doubled fundamental with a wavelength of 800~nm, corresponding to a photon energy of 6.2~eV of the quadrupled beam.

\section{Results and Discussion}

\subsection{Occupied electronic states\label{sect:occupied}}

\begin{figure*}[htbp]
\includegraphics{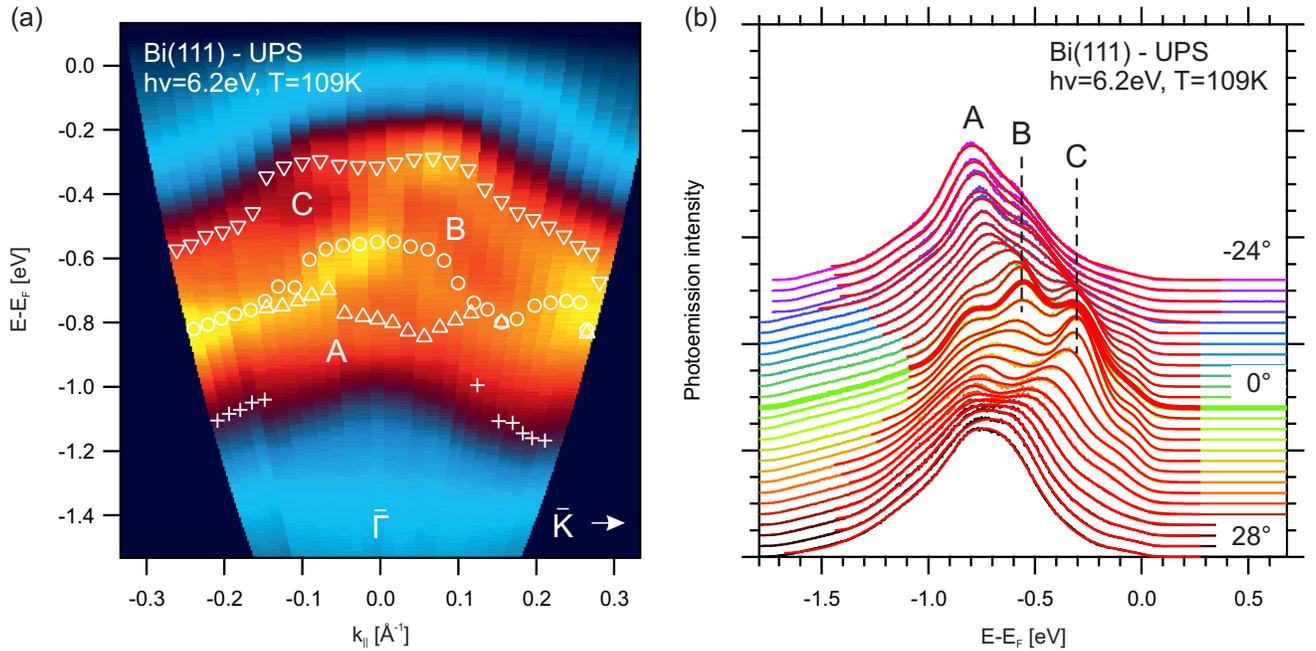}
\caption{(a) ARPES measurement along the $\bar K'\bar\Gamma\bar K$-line depicted in a false-color plot. The fitted peak maxima are indicated with markers for each spectrum. At higher $k_{\parallel}$, peaks A and B overlap which is why one fit component causes an artefact (cross markers). Note that the $\bar\Gamma\bar K$ and $\bar\Gamma\bar K'$ directions are not distinguishable in our experiment.  (b) Single UPS spectra [same data as in (a)] together with the respective fits (see text).\label{fig:Fig2-UPS}}
\end{figure*}

Using the twice frequency-doubled 800~nm fundamental of our laser system, we conducted UPS measurements with a photon energy of 6.2~eV (see Fig. \ref{fig:Fig2-UPS}).
We observe three peaks in the spectra (labeled A, B and C), all of which disperse along the $\bar K'\bar\Gamma\bar K$-line of the SBZ. While the low intensity of A makes it difficult to follow its dispersion in great detail, features B and C generally disperse in a hole-like manner, i.e. with a negative effective mass around the $\bar\Gamma$-point. Note that the normal emission angle and thus the origin of the $k_{\parallel}$ axis was determined from measurements of the delocalized image potential state (see below) and not from the UPS features.
We do not observe any features close to the Fermi edge which could be associated with the spin-split surface band that leads to an electron pocket at $\bar\Gamma$ and six radial hole pockets.\cite{{Ast2001},{Koroteev2004}} One reason for their absence in our spectra is the fact that we do not probe the $\bar\Gamma\bar M$-direction of the SBZ in which the hole pockets are oriented. Furthermore the surface band is only observed in the projected band gap of the bulk \emph{p}-bands of Bi while at the $\bar\Gamma$-point the photoemission intensity is drastically reduced due to the resonance character of this feature associated with the surface.\cite{{Ast2003},{Koroteev2004}}
Peaks B and C clearly show an asymmetry in their intensities with respect to $\bar\Gamma$ which could be a consequence of the threefold symmetric Brillouin zone in which the $\bar\Gamma\bar K$- and $\bar\Gamma\bar K'$-lines are not equivalent (see Fig. \ref{fig:Fig1-structure}a). While both bulk bands and surface states are threefold symmetric, states associated with the first bilayer have a sixfold symmetry, such as the spin-split surface band.\cite{Ast2003}\\
The UPS spectra in Fig. \ref{fig:Fig2-UPS}b were fitted with three independent Gaussian peak profiles on a linear background which was cut off by a Fermi function. The resulting peak positions are displayed in Fig. \ref{fig:Fig2-UPS}a in the false-color plot. Since peak A is less intense than B and C and since it overlaps with B at higher angles, its behavior can not be determined beyond approximately $k_{\parallel}=\pm0.1~{\rm{\AA}}^{-1}$. It was still kept in the fitting function for the sake of consistency.
While state C generally shows a hole-like dispersion, it deviates from a parabolic curve at the $\bar\Gamma$-point where a minimum with electron-like dispersion is found between two extrema at $\pm0.08~{\rm{\AA}}^{-1}$. In contrast to previous ARPES experiments conducted along the $\bar\Gamma\bar K$-line\cite{{Thomas1999},{Patthey1994}} we used a very low photon energy and recorded spectra with relatively high angular resolution around the center of the SBZ. Our quite unusual experimental conditions for ARPES thus provide high momentum resolution at small $k_{\parallel}$ and in fact allowed the first observation of these two distinct maxima in the present study. However, a similar behavior has been reported along the $\bar\Gamma\bar M$ direction and the state was found to have threefold symmetry\cite{Ast2003} which is in agreement with the observed asymmetric intensity distribution in our experiment. At the center of the SBZ we find a binding energy of $E_{\rm{C}}=-0.31\pm0.03~{\rm{eV}}$ which complies well with the surface resonance peak reported in the literature.\cite{{Jezequel1986},{Jezequel1997},{Thomas1999},{Hengsberger2000},{Tanaka1999},{Ast2004},{Patthey1994},{Ast2002},{Ast2003},{Ohtsubo2012}}\\
The binding energy of feature B amounts to $E_{\rm{B}}=-0.55\pm0.03~{\rm{eV}}$ and peak A has a slightly larger binding energy of $E_{\rm{A}}=-0.78\pm0.05~{\rm{eV}}$ at $\bar\Gamma$. Both states seem to be degenerate at higher angles and might be due to the occupied \emph{p}-bands which have been reported to lie in this region\cite{{Jezequel1986},{Jezequel1997},{Thomas1999},{Tanaka1999},{Ast2004}} or another surface resonance which was found at $-0.6$~eV.\cite{Ast2003} The assignment to a \emph{p}-band by comparison to the cited studies is not straightforward due to the different photon energies which probe different points along the $\Gamma T$-line.

\subsection{Unoccupied electronic states}

\begin{figure*}[htbp]
\includegraphics{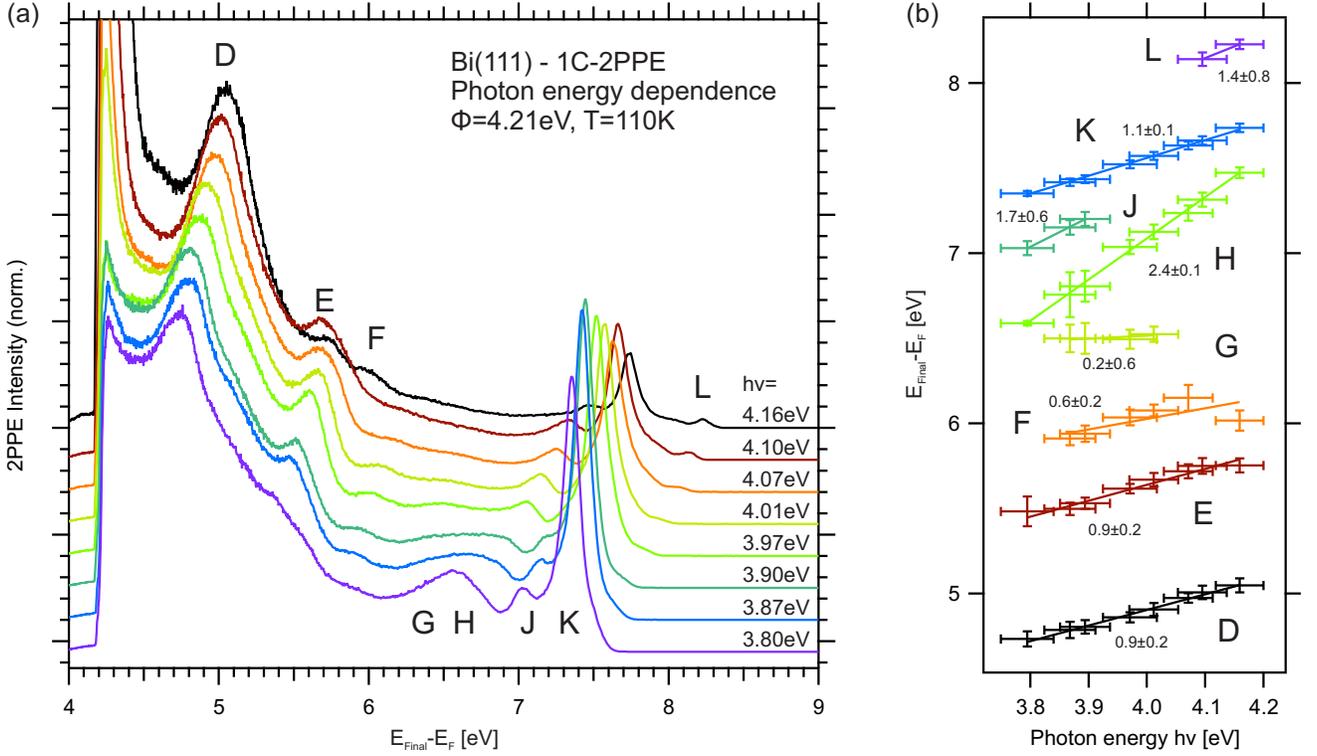}
\caption{(a) 1C-2PPE spectra recorded with different photon energies $h\nu$. The spectra were normalized to the intensity of peak D. (b) Peak shift of the spectral features with varying photon energies. The respective slopes are given in the figure.\label{fig:Fig3-hvDep}}
\end{figure*}

After studying the occupied electronic structure with a one-photon process, we used two-photon photoemission in order to elucidate the energetic position and dispersion of unoccupied states above the Fermi level. Fig. \ref{fig:Fig3-hvDep} shows a series of 1C-2PPE spectra recorded at various photon energies in the UV regime. The spectra are displayed with respect to the final state energy, thus the position of the low energy cut-off directly yields the work function of Bi(111) which has been determined from several different measurements to be $\Phi=4.23\pm0.02~{\rm{eV}}$. This value is compatible with previous experiments.\cite{Muntwiler2008}
A total of eight features are observed in the 2PPE spectra, labeled D-L. The photon energy was varied from 3.80~eV to 4.16~eV which causes a shift of almost all peaks, as depicted in Fig. \ref{fig:Fig3-hvDep}b. As described in sect. \ref{sect:experimental}, in this one-color experiment, a slope of one indicates an unoccupied electronic state whereas a shift with twice the photon energy is characteristic for an occupied state. A peak which does not show a change in peak position upon variation of the photon energy can be assigned to a final state which lies above the vacuum level of the sample. The energies given here are however to be considered with care since, as discussed above (see sect. \ref{sect:experimental}), the dispersion perpendicular to the surface can alter the slope. Peaks D, E and F all shift in a similar manner, namely with a slope of one, although peak F shows a lower energy in the spectrum with highest photon energy, which leads to a slope of less than one and might be due to dispersion along the $\Gamma T$-line or the occurrence of a resonance with an occupied state. These peaks can therefore be assigned to unoccupied states with binding energies of $E_{\rm{D}}=0.91\pm0.07~{\rm{eV}}$, $E_{\rm{E}}=1.64\pm0.06~{\rm{eV}}$ and $E_{\rm{F}}=2.06\pm0.09~{\rm{eV}}$ relative to the Fermi level, respectively. Feature G is relatively weak, does not show a variation of the peak position and overlaps with peak H in the lowest photon energy spectrum. Hence it can be assigned to a final state lying at $E_{\rm{G}}=6.50\pm0.09~{\rm{eV}}$. On the other hand, peaks H and J shift with twice the photon energy and J shifts through peak K at higher photon energies. This behavior would suggest those two peaks to originate from occupied states lying at $E_{\rm{H}}=-0.89\pm0.08~{\rm{eV}}$ and $E_{\rm{J}}=-0.58\pm0.08~{\rm{eV}}$, respectively. Peak K is the dominant feature near the Fermi edge of the spectrum and shifts with a slope of one. From its energetic position of $E_{\rm{K}}=3.56\pm0.05~{\rm{eV}}$ it can be assigned to the first ($n=1$) image potential state (IPS) which has been reported to be located at 3.57~eV.\cite{Muntwiler2008} Peak L can only be observed in two spectra but assuming an unoccupied state (slope one), the binding energy of $E_{\rm{L}}=4.06\pm0.06~{\rm{eV}}$ corresponds very well to the $n=2$ IPS at 4.05~eV.\cite{Muntwiler2008} Since this evaluation scheme is strictly only applicable to states localized at the surface, i.e. without dispersion along the surface normal, we need further measurements characterizing the observed states so that a more detailed comparison with literature values can confirm our conclusions and yield an unambiguous assignment.

\begin{figure}[htbp]
\includegraphics{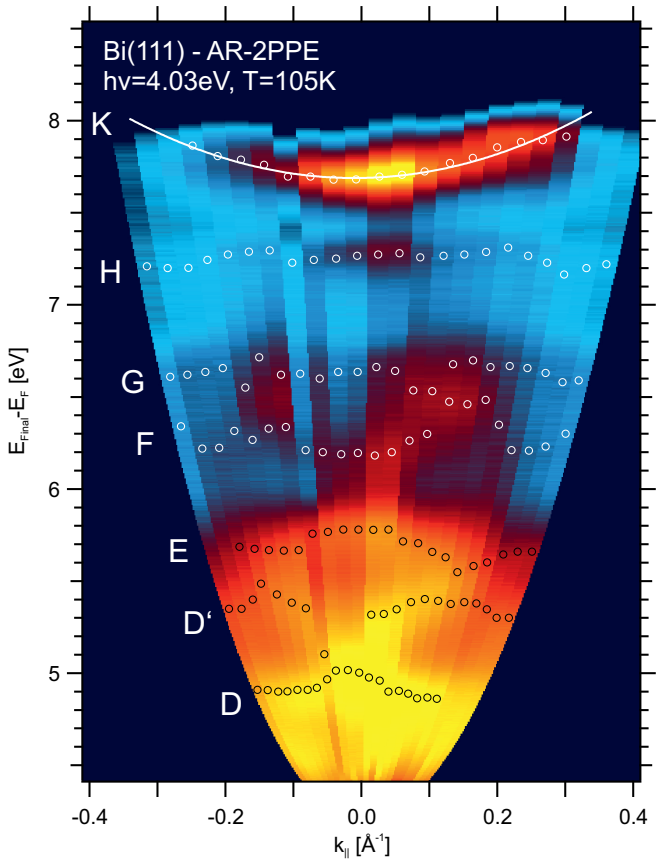}
\caption{Angle-resolved 1C-2PPE measurement at a photon energy of $h\nu=4.03~{\rm{eV}}$. The 2PPE intensity is depicted on a logarithmic color scale in this false-color plot because of the large differences in intensity. The position of the different peaks are indicated with markers. The measurement was performed along the $\bar K'\bar\Gamma\bar K$-line of the SBZ.\label{fig:Fig4-DispUV}}
\end{figure}

One very important property of an electronic state is the dispersion behavior which we investigated using angle-resolved 1C-2PPE (see Fig. \ref{fig:Fig4-DispUV}) along the $\bar K'\bar\Gamma\bar K$-line. For this experiment, the photon energy of $h\nu=4.03~{\rm{eV}}$ was chosen considering the results of the photon energy dependent experiment to obtain dispersion information about as many electronic states as possible. However, since no photon energy which is smaller than the work function, yields spectra that contain a strong signal of all the peaks observed in Fig. \ref{fig:Fig3-hvDep}, peak J is submerged under K in this dispersion measurement. In the low energy range of the spectrum we find that peak D is split into two peaks (labeled D and D') at higher emission angles. While D exhibits a hole-like dispersion, D' disperses to higher binding energies and reaches peak E at around $\pm0.15~{\rm{\AA}}^{-1}$. E also disperses in a hole-like manner around $\bar\Gamma$. At higher $k_{\parallel}$, D' and E either cross each other or E disperses to higher energy while D' turns back toward the Fermi level again. At higher energies, peaks F and G show a similar behavior as D' and E. F disperses in an electron-like fashion around the center of the Brillouin zone, then reaches G at approximately $\pm0.15~{\rm{\AA}}^{-1}$ and either crosses it or both peaks turn around at this point. Note that at the point where both peaks come closest to each other, the border of the Brillouin zone is not reached, as the $\bar K$-point corresponds to $k_{\parallel}=0.9~{\rm{\AA}}^{-1}$. Although peak H can not be observed for large emission angles in the false-color plot, a distinct peak is observable in the spectra and its position is indicated by markers for clarity. The peak does not show a significant dispersion. In the high energy range of the spectrum, peak K shows a parabolic dispersion which can be described by a free electron dispersion with an effective mass $m_{\rm{eff}}$ and the minimum binding energy $E_{0}$ according to the following equation.
\begin{equation}
E(k_{\parallel})=E_{0}+\frac{\hbar^2 k_{\parallel}^2}{2m_{\rm{eff}}}
\end{equation}
Fig. \ref{fig:Fig4-DispUV} shows a corresponding fit and the effective mass deduced from this fit is $m_{\rm{eff}}=1.3\pm0.2m_{\rm{e}}$.

\begin{figure}[htbp]
\includegraphics{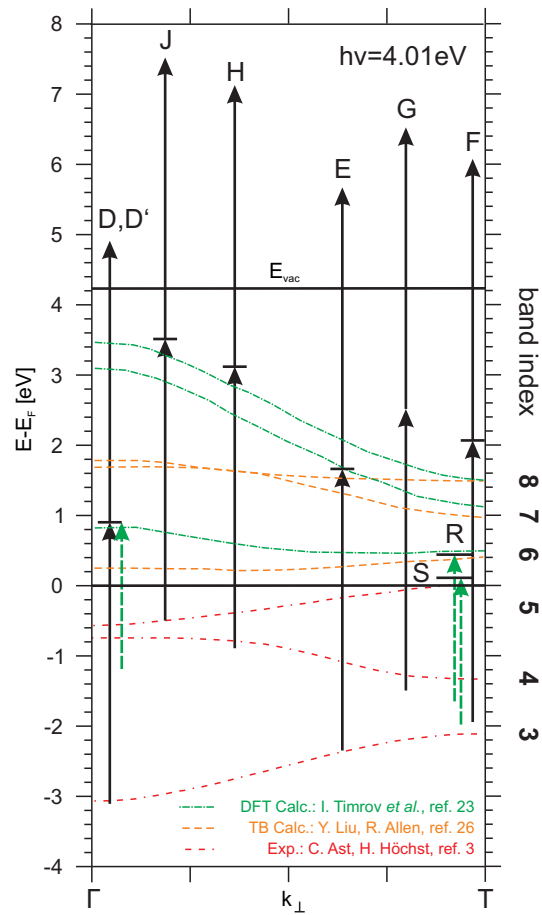}
\caption{Electronic structure along the $\Gamma T$-line and transitions observed in 2PPE experiments presented here. The black arrows indicate photon-induced transitions based on the peaks in the spectrum recorded with $h\nu=4.01~{\rm{eV}}$ in Fig. \ref{fig:Fig3-hvDep}. The position of the transitions along the $k_{\perp}$ direction is based on the coincidence with initial states. Experimental results\cite{Ast2004} on the dispersion of the occupied \emph{p}-bands are shown in the diagram as well as calculated unoccupied \emph{p}-bands.\cite{{Liu1995},{Timrov2012}} The green, dashed arrows indicate alternate population pathways which are relevant in the 2C-2PPE experiments discussed below.\label{fig:Fig5-EnergyDiagram}}
\end{figure}

\begin{figure}[htbp]
\includegraphics{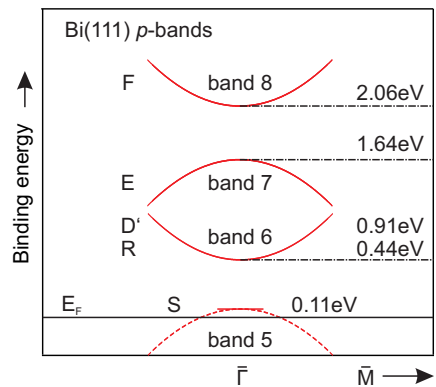}
\caption{Simplified scheme of the band structure above the Fermi level. The three unoccupied $6p$-bands (bands 6-8) are shown as parabolas which is an oversimplified picture. However, the kind of dispersion, i.e. to higher or lower energies, respectively, is indicated in the figure. The dispersion of the hole pocket can not be followed except in a small window around the $\bar\Gamma$-point.\label{fig:Fig6-SchematicBandDiagram}}
\end{figure}

Combining the photon energy dependent measurements in Fig. \ref{fig:Fig3-hvDep}, which might be perturbed by dispersion perpendicular to the surface, with angle-resolved dispersion measurements (Fig. \ref{fig:Fig4-DispUV}) and previously published experiments and calculations we can assign the features in the 2PPE spectra as follows: states K and L are easily assigned to the first two image potential states due to the excellent agreement of their binding energies with a previous study\cite{Muntwiler2008} and the free-electron-like dispersion of K.
For identification of the other features observed in the 2PPE spectra, we have to consider further information from the literature, i.e. a measurement of the three occupied \emph{p}-bands along the $\Gamma T$-line by Ast and H\"ochst\cite{Ast2004} and two different theoretical studies of the occupied and unoccupied bulk band structure.\cite{{Liu1995},{Timrov2012}} Fig. \ref{fig:Fig5-EnergyDiagram} gives an overview of the band dispersion of the bulk \emph{p}-bands along the $\Gamma T$-line. From the observed peak positions in 2PPE ($h\nu=4.01~{\rm{eV}}$, Fig. \ref{fig:Fig3-hvDep}a) we can conclude to the energy of the intermediate and initial states of the respective 2PPE process but we can not determine the momentum perpendicular to the surface experimentally and we are unable to state immediately from the experiments whether a peak arises from an excitation with two photons (i.e. an occupied state excited via a virtual intermediate state) or from an electronic transition that involves unoccupied bands. However, since the initial state energy must coincide with an occupied state, we can estimate qualitatively the location of a transition in reciprocal space and, by carefully analyzing the behavior of the peak (dispersion $k_{\parallel}$ and slope in the photon energy dependent experiment), conclude to the existence and properties of the unoccupied bands. In the following we will adapt the notation introduced by Ast and H\"ochst\cite{Ast2004} which starts with the $6s$-bands (bands 1 and 2) while the occupied $6p$-bands are labeled 3-5 and the unoccupied bands are consequently named 6-8.\\
In the 2PPE spectrum at normal emission (Fig. \ref{fig:Fig3-hvDep}a), the two features D and D' form a single peak at 4.9~eV and, after subtraction of twice the photon energy (see Fig. \ref{fig:Fig5-EnergyDiagram}), it is clear that this final state can only be populated from band 3. Since the peak splits in two features with different dispersion parallel to the surface, one can conclude that not only the occupied band can contribute to the feature in the spectrum. The behavior of band 3 has been determined experimentally to disperse to lower energies\cite{Thomas1999} which is supported by the calculated dispersion of this band along both the $\Gamma K$- and $TW$-lines.\cite{Liu1995} We can hence conclude that while peak D must stem from this occupied band 3, the other peak D' with its electron-like dispersion originates from the unoccupied band 6 lying at 0.9~eV and close to the $\Gamma$-point (see Fig. \ref{fig:Fig5-EnergyDiagram}), in good agreement with the calculation by Timrov \emph{et al}.\cite{Timrov2012} The fact that features D and D' shift with 0.9 times the photon energy (Fig. \ref{fig:Fig3-hvDep}b) indicates that band 6 does not show a strong dispersion along the surface normal close to $\Gamma$.
The final state corresponding to peak E can also, energetically, only be populated from band 3. However, the resulting position in the Brillouin zone does not coincide with the high symmetry points $\Gamma$ or $T$ at which the density of states is highest for this band. We conclude that the relatively high spectral weight of E is thus a consequence of a transition via a real intermediate state, i.e. the unoccupied $p$-band 7 at 1.6~eV, again in excellent agreement with Timrov \emph{et al}.\cite{Timrov2012} This band shows a hole-like dispersion with an effective mass higher than that of band 6 and again the slope of $0.9\pm0.1$ in the photon energy dependent measurement shows that the perpendicular dispersion of this band is not high enough to alter the peak shift significantly from one.
Peak F must also have band 3 as its initial state, but the transition occurs near the $T$-point of the Brillouin zone. Due to the dispersion along the surface to higher energies, which does not comply with the dispersion of band 3,\cite{{Thomas1999},{Liu1995}} and because the peak shift along with the photon energy should be closer to two (instead of 0.6) for an excitation of this band (which is rather flat at the $T$-point) via a virtual intermediate state, we conclude that also in this case an unoccupied $p$-band is involved in the 2PPE process, i.e. band 8. Here, the agreement with the calculations is not as good as in the previous two cases but it should be emphasized that the determination of the perpendicular wave vector $k_{\perp}$ is a rather rough estimate.
Feature G which only slightly shifts along with the varied photon energy (slope of 0.2) must be populated from the second $p$-band, named band 4, and in the vicinity of the $T$-point. We have already learned that in this area of the Brillouin zone, the perpendicular dispersion of both bands 7 and 8 are not sufficient to alter the peak shift in a photon energy dependent measurement from one to 0.2 which however would be necessary if an unoccupied band were involved in this particular 2PPE process. Instead we conclude that a two-photon excitation via a virtual intermediate state occurs from band 4 which disperses enough midway between $\Gamma$ and $T$ such that the final state G can be populated for various excitation photon energies from different points along the $\Gamma T$-line.
Features H and J lie at considerably higher final state energies which makes it more difficult to assign their initial states to either band 4 or 5---or to one of the surface resonances located at $-0.4$~eV\cite{Ast2004} or $-0.6$~eV.\cite{Ast2003} Peak H has a large slope in the photon energy dependence ($2.4\pm0.1$) and does not show a strong dispersion parallel to the surface. In contrast, the surface resonances show a stronger hole-like dispersion (see sect. \ref{sect:occupied} and other photoemission experiments\cite{{Ast2003},{Ast2002},{Patthey1994}}), as do the bulk bands 4 and 5.\cite{Thomas1999} We can thus conclude that this dispersion behavior can be attributed to the intermediate state of this transition, i.e. band 7 or 8. In fact, the moderate, hole-like dispersion resembles more that of band 7 than 8 (see peaks E (7) and F (8) in Fig. \ref{fig:Fig4-DispUV}). The large slope of 2.4 even supports this interpretation of the real intermediate state since the calculation by Timrov \emph{et al}.\cite{Timrov2012} predicts a steep perpendicular dispersion of bands 7 and 8 in the vicinity of the $\Gamma$-point. Although the dispersion of peak J can not be determined since it is submerged under the IPS in the angle-resolved measurement, we can assume a similar scheme as for peak H, involving band 5 or one of the surface resonances and either band 7 or 8 as intermediate states. However, since the slope is also compatible with two, a simple two-photon excitation from a surface resonance could also result in this peak. The energetic position at $\bar\Gamma$ and the dispersion of the three unoccupied $p$-bands along the $\bar\Gamma\bar M$-line of the SBZ is schematically depicted in Fig. \ref{fig:Fig6-SchematicBandDiagram}.
Considering the results presented here it seems that the density functional theory calculations by Timrov \emph{et al}.\cite{Timrov2012} describe the positions of the unoccupied band better than the tight-binding calculations by Liu and Allen.\cite{Liu1995}

\begin{figure}[htbp]
\includegraphics{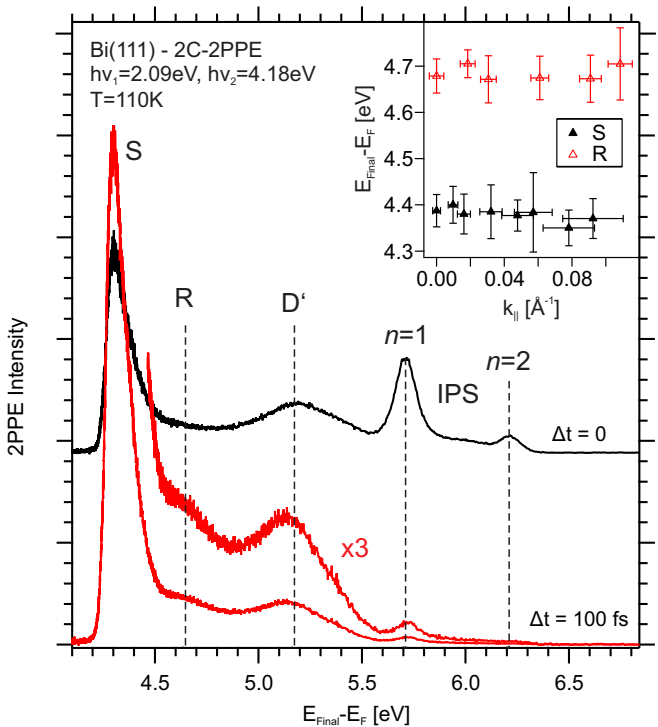}
\caption{Two-color 2PPE spectra without pump-probe delay (black) and with the probe pulse (UV, $h\nu=4.18~{\rm{eV}}$) delayed by 100~fs (red). Unlike the other features, the IPS are pumped with the UV light and therefore their intensity is strongly reduced in the delayed measurement. Peaks R and S on the other hand gain intensity at higher pump-probe delays. The inset shows the results of an angle-resolved 2PPE measurement at a delay of 300~fs.\label{fig:Fig7-RS}}
\end{figure}

While all the results so far were obtained using one-color 2PPE, other electronic transitions are accessible in two-color 2PPE. Fig. \ref{fig:Fig7-RS} shows 2C-2PPE spectra using visible light as pump and its second harmonic as probe beams, respectively. At time-zero, i.e. when there is no delay between the pump and the probe pulse ($\Delta t=0$), four features are found in the spectrum. Unlike in 1C-2PPE, one can conclude to the origin of the spectral features from their behavior as the pump-probe delay is varied, if they possess a lifetime. For example, the two features at $E_{\rm{Final}}-E_{\rm{F}}=5.7~{\rm{eV}}$ and $E_{\rm{Final}}-E_{\rm{F}}=6.2~{\rm{eV}}$, respectively, exhibit a strongly reduced intensity in the spectrum in which the UV pulse arrives after the visible pulse. This is different when the visible pulses are delayed with respect to the UV by 100~fs (not shown here). Considering the energetic position of the peaks, we thus conclude that these two features arise from the two lowest IPS. In the same manner we can identify the other peaks as being probed with the UV pulses.
Another prominent feature is peak D' at $E_{\rm{Final}}-E_{\rm{F}}=5.2~{\rm{eV}}$. This state arises from the unoccupied \emph{p}-band 6. Note that here, the intermediate state is not pumped from band 3 but from band 4 (see Fig. \ref{fig:Fig5-EnergyDiagram}), therefore the contributions from what was labeled peak D above are not present in this spectrum. This is the reason why D' seems slightly shifted to higher energies compared to the 1C-2PPE spectra.
At lower energies, two additional features labeled R and S are observed at binding energies of $E_{\rm{R}}=0.44\pm0.02~{\rm{eV}}$ and $E_{\rm{S}}=0.11\pm0.03~{\rm{eV}}$, respectively, with respect to the Fermi level. The energetic positions are marked in Fig. \ref{fig:Fig5-EnergyDiagram} near the $T$-point, where the pump excitation with the visible photons is possible from bands 3 and 4. Feature R coincides nicely with the calculated position of the lowest-lying $p$-band 6 at the $T$-point. Note that band 6 occurs twice in the 2C-2PPE spectrum, but it is probed in different points in reciprocal space. Feature S lies even closer to the Fermi level, where the hole pocket of the occupied bulk band 5 is located.\cite{{Ast2001},{Golin1968}} Another unoccupied state so close to the Fermi level is the unoccupied part of the spin-split surface state which however is occupied directly at $\bar\Gamma$.\cite{Koroteev2004} Angle-resolved 2PPE measurements yield further information on these states R and S (see inset of Fig. \ref{fig:Fig7-RS}). Due to the small kinetic energy, the momentum parallel to the surface which can be probed is relatively low but in the observed range, neither state exhibits a dispersion. While this non-dispersive behavior matches the expectations on band 6 (to which we have assigned state R) based on band structure calculations,\cite{{Timrov2012},{Liu1995},{Golin1968}} a stronger dispersion would be expected for S because the Fermi surface of the bulk hole pocket only extends to less than $k_{\parallel}=0.1~{\rm{\AA}}^{-1}$, at least along the $\bar\Gamma\bar M$-line of the SBZ,\cite{{Ast2003},{Ast2001}} and the spin-split surface state is only unoccupied for wave vectors larger than $0.05~{\rm{\AA}}^{-1}$ along the $\bar\Gamma\bar K$-line.\cite{Ast2003}

\section{Conclusions}

Using angle-resolved UPS and 2PPE we have investigated both the occupied and unoccupied electronic structure of Bi along the $\bar\Gamma\bar K$-line of the SBZ.
In UPS, we observe a surface resonance at $-0.31$~eV which has a peculiar (generally hole-like) dispersion with two distinct maxima at $\pm0.08~{\rm{\AA}}^{-1}$. We also find two additional features in the UPS spectra which are associated with the occupied $p$-bands or possibly another surface resonance.
2PPE allows us to induce transitions between the occupied and unoccupied $p$-bands and to directly probe the band structure above the Fermi level. We found spectral features from all three unoccupied $p$-bands and were able to probe their dispersion in the $\bar\Gamma\bar K$-direction. The measured energetic positions are in good agreement with calculations by Timrov \emph{et al}.\cite{Timrov2012} Additionally we observe a final state above the vacuum level of the sample as well as signatures from two resonant excitations between occupied and unoccupied $p$-bands (from band 4 to 7 and from 5 to 8, respectively). Close to the Fermi level, at 0.11~eV, we observe a state which is attributed to the hole pocket formed by the highest occupied $p$-band.
Our experimental study of the unoccupied bands complements the large variety of ARPES studies of the occupied bands and surface states and hence provides an extended picture of the entire band structure of the semi-metallic Bi(111) surface.

\section*{Acknowledgments}
Funding through the Collaborative Research Center Sfb658 of the German Research Foundation (DFG) is gratefully acknowledged.


\begin{thebibliography}{34}%
\makeatletter
\providecommand \@ifxundefined [1]{%
 \@ifx{#1\undefined}
}%
\providecommand \@ifnum [1]{%
 \ifnum #1\expandafter \@firstoftwo
 \else \expandafter \@secondoftwo
 \fi
}%
\providecommand \@ifx [1]{%
 \ifx #1\expandafter \@firstoftwo
 \else \expandafter \@secondoftwo
 \fi
}%
\providecommand \natexlab [1]{#1}%
\providecommand \enquote  [1]{``#1''}%
\providecommand \bibnamefont  [1]{#1}%
\providecommand \bibfnamefont [1]{#1}%
\providecommand \citenamefont [1]{#1}%
\providecommand \href@noop [0]{\@secondoftwo}%
\providecommand \href [0]{\begingroup \@sanitize@url \@href}%
\providecommand \@href[1]{\@@startlink{#1}\@@href}%
\providecommand \@@href[1]{\endgroup#1\@@endlink}%
\providecommand \@sanitize@url [0]{\catcode `\\12\catcode `\$12\catcode
  `\&12\catcode `\#12\catcode `\^12\catcode `\_12\catcode `\%12\relax}%
\providecommand \@@startlink[1]{}%
\providecommand \@@endlink[0]{}%
\providecommand \url  [0]{\begingroup\@sanitize@url \@url }%
\providecommand \@url [1]{\endgroup\@href {#1}{\urlprefix }}%
\providecommand \urlprefix  [0]{URL }%
\providecommand \Eprint [0]{\href }%
\@ifxundefined \urlstyle {%
  \providecommand \doi  [0]{\begingroup \@sanitize@url \@doi}%
  \providecommand \@doi [1]{\endgroup \@@startlink {\doibase
  #1}doi:\discretionary {}{}{}#1\@@endlink }%
}{%
  \providecommand \doi  [0]{doi:\discretionary{}{}{}\begingroup
  \urlstyle{rm}\Url }%
}%
\providecommand \doibase [0]{http://dx.doi.org/}%
\providecommand \Doi [0]{\begingroup \@sanitize@url \@Doi }%
\providecommand \@Doi  [1]{\endgroup\@@startlink{\doibase#1}\@@Doi}%
\providecommand \@@Doi [1]{#1\@@endlink}%
\providecommand \selectlanguage [0]{\@gobble}%
\providecommand \bibinfo  [0]{\@secondoftwo}%
\providecommand \bibfield  [0]{\@secondoftwo}%
\providecommand \translation [1]{[#1]}%
\providecommand \BibitemOpen [0]{}%
\providecommand \bibitemStop [0]{}%
\providecommand \bibitemNoStop [0]{.\EOS\space}%
\providecommand \EOS [0]{\spacefactor3000\relax}%
\providecommand \BibitemShut  [1]{\csname bibitem#1\endcsname}%
\bibitem [{\citenamefont {Hofmann}(2006)}]{Hofmann2006}%
  \BibitemOpen
  \bibfield  {author} {\bibinfo {author} {\bibfnamefont {P.}~\bibnamefont
  {Hofmann}},\ }\Doi {10.1016/j.progsurf.2006.03.001} {\bibfield  {journal}
  {\bibinfo  {journal} {Prog. Surf. Sci.},\ }\textbf {\bibinfo {volume} {81}},\
  \bibinfo {pages} {191} (\bibinfo {year} {2006})}\BibitemShut {NoStop}%
\bibitem [{\citenamefont {Ast}\ and\ \citenamefont
  {H\"{o}chst}(2002)}]{Ast2002}%
  \BibitemOpen
  \bibfield  {author} {\bibinfo {author} {\bibfnamefont {C.~R.}\ \bibnamefont
  {Ast}}\ and\ \bibinfo {author} {\bibfnamefont {H.}~\bibnamefont
  {H\"{o}chst}},\ }\Doi {10.1103/PhysRevB.66.125103} {\bibfield  {journal}
  {\bibinfo  {journal} {Phys. Rev. B},\ }\textbf {\bibinfo {volume} {66}},\
  \bibinfo {pages} {125103} (\bibinfo {year} {2002})}\BibitemShut {NoStop}%
\bibitem [{\citenamefont {Ast}\ and\ \citenamefont
  {H\"{o}chst}(2004)}]{Ast2004}%
  \BibitemOpen
  \bibfield  {author} {\bibinfo {author} {\bibfnamefont {C.~R.}\ \bibnamefont
  {Ast}}\ and\ \bibinfo {author} {\bibfnamefont {H.}~\bibnamefont
  {H\"{o}chst}},\ }\Doi {10.1103/PhysRevB.70.245122} {\bibfield  {journal}
  {\bibinfo  {journal} {Phys. Rev. B},\ }\textbf {\bibinfo {volume} {70}},\
  \bibinfo {pages} {245122} (\bibinfo {year} {2004})}\BibitemShut {NoStop}%
\bibitem [{\citenamefont {Jezequel}\ \emph {et~al.}(1986)\citenamefont
  {Jezequel}, \citenamefont {Petroff}, \citenamefont {Pinchaux},\ and\
  \citenamefont {Yndurain}}]{Jezequel1986}%
  \BibitemOpen
  \bibfield  {author} {\bibinfo {author} {\bibfnamefont {G.}~\bibnamefont
  {Jezequel}}, \bibinfo {author} {\bibfnamefont {Y.}~\bibnamefont {Petroff}},
  \bibinfo {author} {\bibfnamefont {R.}~\bibnamefont {Pinchaux}}, \ and\
  \bibinfo {author} {\bibfnamefont {F.}~\bibnamefont {Yndurain}},\ }\Doi
  {10.1103/PhysRevB.33.4352} {\bibfield  {journal} {\bibinfo  {journal} {Phys.
  Rev. B},\ }\textbf {\bibinfo {volume} {33}},\ \bibinfo {pages} {4352}
  (\bibinfo {year} {1986})}\BibitemShut {NoStop}%
\bibitem [{\citenamefont {Hengsberger}\ \emph {et~al.}(2000)\citenamefont
  {Hengsberger}, \citenamefont {Segovia}, \citenamefont {Garnier},
  \citenamefont {Purdie},\ and\ \citenamefont {Baer}}]{Hengsberger2000}%
  \BibitemOpen
  \bibfield  {author} {\bibinfo {author} {\bibfnamefont {M.}~\bibnamefont
  {Hengsberger}}, \bibinfo {author} {\bibfnamefont {P.}~\bibnamefont
  {Segovia}}, \bibinfo {author} {\bibfnamefont {M.}~\bibnamefont {Garnier}},
  \bibinfo {author} {\bibfnamefont {D.}~\bibnamefont {Purdie}}, \ and\ \bibinfo
  {author} {\bibfnamefont {Y.}~\bibnamefont {Baer}},\ }\Doi
  {10.1007/s100510070097} {\bibfield  {journal} {\bibinfo  {journal} {Eur.
  Phys. J. B},\ }\textbf {\bibinfo {volume} {17}},\ \bibinfo {pages} {603}
  (\bibinfo {year} {2000})}\BibitemShut {NoStop}%
\bibitem [{\citenamefont {Ast}\ and\ \citenamefont
  {H\"{o}chst}(2001)}]{Ast2001}%
  \BibitemOpen
  \bibfield  {author} {\bibinfo {author} {\bibfnamefont {C.~R.}\ \bibnamefont
  {Ast}}\ and\ \bibinfo {author} {\bibfnamefont {H.}~\bibnamefont
  {H\"{o}chst}},\ }\Doi {10.1103/PhysRevLett.87.177602} {\bibfield  {journal}
  {\bibinfo  {journal} {Phys. Rev. Lett.},\ }\textbf {\bibinfo {volume} {87}},\
  \bibinfo {pages} {177602} (\bibinfo {year} {2001})}\BibitemShut {NoStop}%
\bibitem [{\citenamefont {Koroteev}\ \emph {et~al.}(2004)\citenamefont
  {Koroteev}, \citenamefont {Bihlmayer}, \citenamefont {Gayone}, \citenamefont
  {Chulkov}, \citenamefont {Bl\"{u}gel}, \citenamefont {Echenique},\ and\
  \citenamefont {Hofmann}}]{Koroteev2004}%
  \BibitemOpen
  \bibfield  {author} {\bibinfo {author} {\bibfnamefont {Y.~M.}\ \bibnamefont
  {Koroteev}}, \bibinfo {author} {\bibfnamefont {G.}~\bibnamefont {Bihlmayer}},
  \bibinfo {author} {\bibfnamefont {J.~E.}\ \bibnamefont {Gayone}}, \bibinfo
  {author} {\bibfnamefont {E.~V.}\ \bibnamefont {Chulkov}}, \bibinfo {author}
  {\bibfnamefont {S.}~\bibnamefont {Bl\"{u}gel}}, \bibinfo {author}
  {\bibfnamefont {P.~M.}\ \bibnamefont {Echenique}}, \ and\ \bibinfo {author}
  {\bibfnamefont {P.}~\bibnamefont {Hofmann}},\ }\Doi
  {10.1103/PhysRevLett.93.046403} {\bibfield  {journal} {\bibinfo  {journal}
  {Phys. Rev. Lett.},\ }\textbf {\bibinfo {volume} {93}},\ \bibinfo {pages}
  {046403} (\bibinfo {year} {2004})}\BibitemShut {NoStop}%
\bibitem [{\citenamefont {Ast}\ and\ \citenamefont
  {H\"{o}chst}(2003){\natexlab{a}}}]{Ast2003}%
  \BibitemOpen
  \bibfield  {author} {\bibinfo {author} {\bibfnamefont {C.~R.}\ \bibnamefont
  {Ast}}\ and\ \bibinfo {author} {\bibfnamefont {H.}~\bibnamefont
  {H\"{o}chst}},\ }\Doi {10.1103/PhysRevB.67.113102} {\bibfield  {journal}
  {\bibinfo  {journal} {Phys. Rev. B},\ }\textbf {\bibinfo {volume} {67}},\
  \bibinfo {pages} {113102} (\bibinfo {year} {2003}{\natexlab{a}})}\BibitemShut
  {NoStop}%
\bibitem [{\citenamefont {Ohtsubo}\ \emph {et~al.}(2012)\citenamefont
  {Ohtsubo}, \citenamefont {Mauchain}, \citenamefont {Faure}, \citenamefont
  {Papalazarou}, \citenamefont {Marsi}, \citenamefont {F\'{e}vre},
  \citenamefont {Bertran}, \citenamefont {Taleb-Ibrahimi},\ and\ \citenamefont
  {Perfetti}}]{Ohtsubo2012}%
  \BibitemOpen
  \bibfield  {author} {\bibinfo {author} {\bibfnamefont {Y.}~\bibnamefont
  {Ohtsubo}}, \bibinfo {author} {\bibfnamefont {J.}~\bibnamefont {Mauchain}},
  \bibinfo {author} {\bibfnamefont {J.}~\bibnamefont {Faure}}, \bibinfo
  {author} {\bibfnamefont {E.}~\bibnamefont {Papalazarou}}, \bibinfo {author}
  {\bibfnamefont {M.}~\bibnamefont {Marsi}}, \bibinfo {author} {\bibfnamefont
  {P.~L.}\ \bibnamefont {F\'{e}vre}}, \bibinfo {author} {\bibfnamefont
  {F.}~\bibnamefont {Bertran}}, \bibinfo {author} {\bibfnamefont
  {A.}~\bibnamefont {Taleb-Ibrahimi}}, \ and\ \bibinfo {author} {\bibfnamefont
  {L.}~\bibnamefont {Perfetti}},\ }\href {http://arxiv.org/abs/1209.1219}
  {\bibfield  {journal} {\bibinfo  {journal} {arXiv},\ \bibinfo {pages}
  {arXiv:1209.1219}} (\bibinfo {year} {2012})}\BibitemShut {NoStop}%
\bibitem [{\citenamefont {Datta}\ and\ \citenamefont {Das}(1990)}]{Datta1990}%
  \BibitemOpen
  \bibfield  {author} {\bibinfo {author} {\bibfnamefont {S.}~\bibnamefont
  {Datta}}\ and\ \bibinfo {author} {\bibfnamefont {B.}~\bibnamefont {Das}},\
  }\Doi {10.1063/1.102730} {\bibfield  {journal} {\bibinfo  {journal} {Appl.
  Phys. Lett.},\ }\textbf {\bibinfo {volume} {56}},\ \bibinfo {pages} {665}
  (\bibinfo {year} {1990})}\BibitemShut {NoStop}%
\bibitem [{\citenamefont {Koga}\ \emph {et~al.}(2002)\citenamefont {Koga},
  \citenamefont {Nitta}, \citenamefont {Takayanagi},\ and\ \citenamefont
  {Datta}}]{Koga2002}%
  \BibitemOpen
  \bibfield  {author} {\bibinfo {author} {\bibfnamefont {T.}~\bibnamefont
  {Koga}}, \bibinfo {author} {\bibfnamefont {J.}~\bibnamefont {Nitta}},
  \bibinfo {author} {\bibfnamefont {H.}~\bibnamefont {Takayanagi}}, \ and\
  \bibinfo {author} {\bibfnamefont {S.}~\bibnamefont {Datta}},\ }\Doi
  {10.1103/PhysRevLett.88.126601} {\bibfield  {journal} {\bibinfo  {journal}
  {Phys. Rev. Lett.},\ }\textbf {\bibinfo {volume} {88}},\ \bibinfo {pages}
  {126601} (\bibinfo {year} {2002})}\BibitemShut {NoStop}%
\bibitem [{\citenamefont {Vossloh}\ \emph {et~al.}(1998)\citenamefont
  {Vossloh}, \citenamefont {Holdenried},\ and\ \citenamefont
  {Micklitz}}]{Vossloh1998}%
  \BibitemOpen
  \bibfield  {author} {\bibinfo {author} {\bibfnamefont {C.}~\bibnamefont
  {Vossloh}}, \bibinfo {author} {\bibfnamefont {M.}~\bibnamefont {Holdenried}},
  \ and\ \bibinfo {author} {\bibfnamefont {H.}~\bibnamefont {Micklitz}},\ }\Doi
  {10.1103/PhysRevB.58.12422} {\bibfield  {journal} {\bibinfo  {journal} {Phys.
  Rev. B},\ }\textbf {\bibinfo {volume} {58}},\ \bibinfo {pages} {12422}
  (\bibinfo {year} {1998})}\BibitemShut {NoStop}%
\bibitem [{\citenamefont {Weitzel}\ and\ \citenamefont
  {Micklitz}(1991)}]{Weitzel1991}%
  \BibitemOpen
  \bibfield  {author} {\bibinfo {author} {\bibfnamefont {B.}~\bibnamefont
  {Weitzel}}\ and\ \bibinfo {author} {\bibfnamefont {H.}~\bibnamefont
  {Micklitz}},\ }\Doi {10.1103/PhysRevLett.66.385} {\bibfield  {journal}
  {\bibinfo  {journal} {Phys. Rev. Lett.},\ }\textbf {\bibinfo {volume} {66}},\
  \bibinfo {pages} {385} (\bibinfo {year} {1991})}\BibitemShut {NoStop}%
\bibitem [{\citenamefont {Patthey}\ \emph {et~al.}(1994)\citenamefont
  {Patthey}, \citenamefont {Schneider},\ and\ \citenamefont
  {Micklitz}}]{Patthey1994}%
  \BibitemOpen
  \bibfield  {author} {\bibinfo {author} {\bibfnamefont {F.}~\bibnamefont
  {Patthey}}, \bibinfo {author} {\bibfnamefont {W.-D.}\ \bibnamefont
  {Schneider}}, \ and\ \bibinfo {author} {\bibfnamefont {H.}~\bibnamefont
  {Micklitz}},\ }\Doi {10.1103/PhysRevB.49.11293} {\bibfield  {journal}
  {\bibinfo  {journal} {Phys. Rev. B},\ }\textbf {\bibinfo {volume} {49}},\
  \bibinfo {pages} {11293} (\bibinfo {year} {1994})}\BibitemShut {NoStop}%
\bibitem [{\citenamefont {Wells}\ \emph {et~al.}(2009)\citenamefont {Wells},
  \citenamefont {Dil}, \citenamefont {Meier}, \citenamefont {Lobo-Checa},
  \citenamefont {Petrov}, \citenamefont {Osterwalder}, \citenamefont {Ugeda},
  \citenamefont {Fernandez-Torrente}, \citenamefont {Pascual}, \citenamefont
  {Rienks}, \citenamefont {Jensen},\ and\ \citenamefont {Hofmann}}]{Wells2009}%
  \BibitemOpen
  \bibfield  {author} {\bibinfo {author} {\bibfnamefont {J.~W.}\ \bibnamefont
  {Wells}}, \bibinfo {author} {\bibfnamefont {J.~H.}\ \bibnamefont {Dil}},
  \bibinfo {author} {\bibfnamefont {F.}~\bibnamefont {Meier}}, \bibinfo
  {author} {\bibfnamefont {J.}~\bibnamefont {Lobo-Checa}}, \bibinfo {author}
  {\bibfnamefont {V.~N.}\ \bibnamefont {Petrov}}, \bibinfo {author}
  {\bibfnamefont {J.}~\bibnamefont {Osterwalder}}, \bibinfo {author}
  {\bibfnamefont {M.~M.}\ \bibnamefont {Ugeda}}, \bibinfo {author}
  {\bibfnamefont {I.}~\bibnamefont {Fernandez-Torrente}}, \bibinfo {author}
  {\bibfnamefont {J.~I.}\ \bibnamefont {Pascual}}, \bibinfo {author}
  {\bibfnamefont {E.~D.~L.}\ \bibnamefont {Rienks}}, \bibinfo {author}
  {\bibfnamefont {M.~F.}\ \bibnamefont {Jensen}}, \ and\ \bibinfo {author}
  {\bibfnamefont {P.}~\bibnamefont {Hofmann}},\ }\Doi
  {10.1103/PhysRevLett.102.096802} {\bibfield  {journal} {\bibinfo  {journal}
  {Phys. Rev. Lett.},\ }\textbf {\bibinfo {volume} {102}},\ \bibinfo {pages}
  {096802} (\bibinfo {year} {2009})}\BibitemShut {NoStop}%
\bibitem [{\citenamefont {Hsieh}\ \emph {et~al.}(2008)\citenamefont {Hsieh},
  \citenamefont {Qian}, \citenamefont {Wray}, \citenamefont {Xia},
  \citenamefont {Hor}, \citenamefont {Cava},\ and\ \citenamefont
  {Hasan}}]{Hsieh2008}%
  \BibitemOpen
  \bibfield  {author} {\bibinfo {author} {\bibfnamefont {D.}~\bibnamefont
  {Hsieh}}, \bibinfo {author} {\bibfnamefont {D.}~\bibnamefont {Qian}},
  \bibinfo {author} {\bibfnamefont {L.}~\bibnamefont {Wray}}, \bibinfo {author}
  {\bibfnamefont {Y.}~\bibnamefont {Xia}}, \bibinfo {author} {\bibfnamefont
  {Y.~S.}\ \bibnamefont {Hor}}, \bibinfo {author} {\bibfnamefont {R.~J.}\
  \bibnamefont {Cava}}, \ and\ \bibinfo {author} {\bibfnamefont {M.~Z.}\
  \bibnamefont {Hasan}},\ }\Doi {10.1038/nature06843} {\bibfield  {journal}
  {\bibinfo  {journal} {Nature},\ }\textbf {\bibinfo {volume} {452}},\ \bibinfo
  {pages} {970} (\bibinfo {year} {2008})}\BibitemShut {NoStop}%
\bibitem [{\citenamefont {Zhang}\ \emph {et~al.}(2009)\citenamefont {Zhang},
  \citenamefont {Liu}, \citenamefont {Qi}, \citenamefont {Dai}, \citenamefont
  {Fang},\ and\ \citenamefont {Zhang}}]{Zhang2009}%
  \BibitemOpen
  \bibfield  {author} {\bibinfo {author} {\bibfnamefont {H.}~\bibnamefont
  {Zhang}}, \bibinfo {author} {\bibfnamefont {C.-X.}\ \bibnamefont {Liu}},
  \bibinfo {author} {\bibfnamefont {X.-L.}\ \bibnamefont {Qi}}, \bibinfo
  {author} {\bibfnamefont {X.}~\bibnamefont {Dai}}, \bibinfo {author}
  {\bibfnamefont {Z.}~\bibnamefont {Fang}}, \ and\ \bibinfo {author}
  {\bibfnamefont {S.-C.}\ \bibnamefont {Zhang}},\ }\Doi {10.1038/nphys1270}
  {\bibfield  {journal} {\bibinfo  {journal} {Nature Phys.},\ }\textbf
  {\bibinfo {volume} {5}},\ \bibinfo {pages} {438} (\bibinfo {year}
  {2009})}\BibitemShut {NoStop}%
\bibitem [{\citenamefont {Jezequel}\ \emph {et~al.}(1997)\citenamefont
  {Jezequel}, \citenamefont {Thomas},\ and\ \citenamefont
  {Pollini}}]{Jezequel1997}%
  \BibitemOpen
  \bibfield  {author} {\bibinfo {author} {\bibfnamefont {G.}~\bibnamefont
  {Jezequel}}, \bibinfo {author} {\bibfnamefont {J.}~\bibnamefont {Thomas}}, \
  and\ \bibinfo {author} {\bibfnamefont {I.}~\bibnamefont {Pollini}},\ }\Doi
  {10.1103/PhysRevB.56.6620} {\bibfield  {journal} {\bibinfo  {journal} {Phys.
  Rev. B},\ }\textbf {\bibinfo {volume} {56}},\ \bibinfo {pages} {6620}
  (\bibinfo {year} {1997})}\BibitemShut {NoStop}%
\bibitem [{\citenamefont {Thomas}\ \emph {et~al.}(1999)\citenamefont {Thomas},
  \citenamefont {Jezequel},\ and\ \citenamefont {Pollini}}]{Thomas1999}%
  \BibitemOpen
  \bibfield  {author} {\bibinfo {author} {\bibfnamefont {J.}~\bibnamefont
  {Thomas}}, \bibinfo {author} {\bibfnamefont {G.}~\bibnamefont {Jezequel}}, \
  and\ \bibinfo {author} {\bibfnamefont {I.}~\bibnamefont {Pollini}},\ }\href
  {http://iopscience.iop.org/0953-8984/11/48/314} {\bibfield  {journal}
  {\bibinfo  {journal} {J. Phys.: Condens. Matter},\ }\textbf {\bibinfo
  {volume} {11}},\ \bibinfo {pages} {9571} (\bibinfo {year}
  {1999})}\BibitemShut {NoStop}%
\bibitem [{\citenamefont {Tanaka}\ \emph {et~al.}(1999)\citenamefont {Tanaka},
  \citenamefont {Hatano}, \citenamefont {Takahashi}, \citenamefont {Sasaki},
  \citenamefont {Suzuki},\ and\ \citenamefont {Sato}}]{Tanaka1999}%
  \BibitemOpen
  \bibfield  {author} {\bibinfo {author} {\bibfnamefont {A.}~\bibnamefont
  {Tanaka}}, \bibinfo {author} {\bibfnamefont {M.}~\bibnamefont {Hatano}},
  \bibinfo {author} {\bibfnamefont {K.}~\bibnamefont {Takahashi}}, \bibinfo
  {author} {\bibfnamefont {H.}~\bibnamefont {Sasaki}}, \bibinfo {author}
  {\bibfnamefont {S.}~\bibnamefont {Suzuki}}, \ and\ \bibinfo {author}
  {\bibfnamefont {S.}~\bibnamefont {Sato}},\ }\Doi
  {10.1016/S0039-6028(99)00088-6} {\bibfield  {journal} {\bibinfo  {journal}
  {Surf. Sci.},\ }\textbf {\bibinfo {volume} {433-435}},\ \bibinfo {pages}
  {647} (\bibinfo {year} {1999})}\BibitemShut {NoStop}%
\bibitem [{\citenamefont {Ast}\ and\ \citenamefont
  {H\"{o}chst}(2003){\natexlab{b}}}]{Ast2003a}%
  \BibitemOpen
  \bibfield  {author} {\bibinfo {author} {\bibfnamefont {C.~R.}\ \bibnamefont
  {Ast}}\ and\ \bibinfo {author} {\bibfnamefont {H.}~\bibnamefont
  {H\"{o}chst}},\ }\Doi {10.1103/PhysRevLett.90.016403} {\bibfield  {journal}
  {\bibinfo  {journal} {Phys. Rev. Lett.},\ }\textbf {\bibinfo {volume} {90}},\
  \bibinfo {pages} {016403} (\bibinfo {year} {2003}{\natexlab{b}})}\BibitemShut
  {NoStop}%
\bibitem [{\citenamefont {Muntwiler}\ and\ \citenamefont
  {Zhu}(2008)}]{Muntwiler2008}%
  \BibitemOpen
  \bibfield  {author} {\bibinfo {author} {\bibfnamefont {M.}~\bibnamefont
  {Muntwiler}}\ and\ \bibinfo {author} {\bibfnamefont {X.-Y.}\ \bibnamefont
  {Zhu}},\ }\Doi {10.1088/1367-2630/10/11/113018} {\bibfield  {journal}
  {\bibinfo  {journal} {New J. Phys.},\ }\textbf {\bibinfo {volume} {10}},\
  \bibinfo {pages} {113018} (\bibinfo {year} {2008})}\BibitemShut {NoStop}%
\bibitem [{\citenamefont {Timrov}\ \emph {et~al.}(2012)\citenamefont {Timrov},
  \citenamefont {Kampfrath}, \citenamefont {Faure}, \citenamefont {Vast},
  \citenamefont {Ast}, \citenamefont {Frischkorn}, \citenamefont {Wolf},
  \citenamefont {Gava},\ and\ \citenamefont {Perfetti}}]{Timrov2012}%
  \BibitemOpen
  \bibfield  {author} {\bibinfo {author} {\bibfnamefont {I.}~\bibnamefont
  {Timrov}}, \bibinfo {author} {\bibfnamefont {T.}~\bibnamefont {Kampfrath}},
  \bibinfo {author} {\bibfnamefont {J.}~\bibnamefont {Faure}}, \bibinfo
  {author} {\bibfnamefont {N.}~\bibnamefont {Vast}}, \bibinfo {author}
  {\bibfnamefont {C.~R.}\ \bibnamefont {Ast}}, \bibinfo {author} {\bibfnamefont
  {C.}~\bibnamefont {Frischkorn}}, \bibinfo {author} {\bibfnamefont
  {M.}~\bibnamefont {Wolf}}, \bibinfo {author} {\bibfnamefont {P.}~\bibnamefont
  {Gava}}, \ and\ \bibinfo {author} {\bibfnamefont {L.}~\bibnamefont
  {Perfetti}},\ }\Doi {10.1103/PhysRevB.85.155139} {\bibfield  {journal}
  {\bibinfo  {journal} {Phys. Rev. B},\ }\textbf {\bibinfo {volume} {85}},\
  \bibinfo {pages} {155139} (\bibinfo {year} {2012})}\BibitemShut {NoStop}%
\bibitem [{\citenamefont {Golin}(1968)}]{Golin1968}%
  \BibitemOpen
  \bibfield  {author} {\bibinfo {author} {\bibfnamefont {S.}~\bibnamefont
  {Golin}},\ }\Doi {10.1103/PhysRev.166.643} {\bibfield  {journal} {\bibinfo
  {journal} {Phys. Rev.},\ }\textbf {\bibinfo {volume} {166}},\ \bibinfo
  {pages} {643} (\bibinfo {year} {1968})}\BibitemShut {NoStop}%
\bibitem [{\citenamefont {Gonze}\ \emph {et~al.}(1990)\citenamefont {Gonze},
  \citenamefont {Michenaud},\ and\ \citenamefont {Vigneron}}]{Gonze1990}%
  \BibitemOpen
  \bibfield  {author} {\bibinfo {author} {\bibfnamefont {X.}~\bibnamefont
  {Gonze}}, \bibinfo {author} {\bibfnamefont {J.-P.}\ \bibnamefont
  {Michenaud}}, \ and\ \bibinfo {author} {\bibfnamefont {J.-P.}\ \bibnamefont
  {Vigneron}},\ }\Doi {10.1103/PhysRevB.41.11827} {\bibfield  {journal}
  {\bibinfo  {journal} {Phys. Rev. B},\ }\textbf {\bibinfo {volume} {41}},\
  \bibinfo {pages} {11827} (\bibinfo {year} {1990})}\BibitemShut {NoStop}%
\bibitem [{\citenamefont {Liu}\ and\ \citenamefont {Allen}(1995)}]{Liu1995}%
  \BibitemOpen
  \bibfield  {author} {\bibinfo {author} {\bibfnamefont {Y.}~\bibnamefont
  {Liu}}\ and\ \bibinfo {author} {\bibfnamefont {R.~E.}\ \bibnamefont
  {Allen}},\ }\Doi {10.1103/PhysRevB.52.1566} {\bibfield  {journal} {\bibinfo
  {journal} {Phys. Rev. B},\ }\textbf {\bibinfo {volume} {52}},\ \bibinfo
  {pages} {1566} (\bibinfo {year} {1995})}\BibitemShut {NoStop}%
\bibitem [{\citenamefont {Weinelt}(2002)}]{Weinelt2002}%
  \BibitemOpen
  \bibfield  {author} {\bibinfo {author} {\bibfnamefont {M.}~\bibnamefont
  {Weinelt}},\ }\href {http://iopscience.iop.org/0953-8984/14/43/202}
  {\bibfield  {journal} {\bibinfo  {journal} {J. Phys.: Condens. Matter},\
  }\textbf {\bibinfo {volume} {14}},\ \bibinfo {pages} {R1099} (\bibinfo {year}
  {2002})}\BibitemShut {NoStop}%
\bibitem [{\citenamefont {Petek}\ and\ \citenamefont
  {Ogawa}(1997)}]{Petek1997}%
  \BibitemOpen
  \bibfield  {author} {\bibinfo {author} {\bibfnamefont {H.}~\bibnamefont
  {Petek}}\ and\ \bibinfo {author} {\bibfnamefont {S.}~\bibnamefont {Ogawa}},\
  }\Doi {10.1016/S0079-6816(98)00002-1} {\bibfield  {journal} {\bibinfo
  {journal} {Prog. Surf. Sci.},\ }\textbf {\bibinfo {volume} {56}},\ \bibinfo
  {pages} {239} (\bibinfo {year} {1997})}\BibitemShut {NoStop}%
\bibitem [{\citenamefont {Zhu}(2004)}]{Zhu2004}%
  \BibitemOpen
  \bibfield  {author} {\bibinfo {author} {\bibfnamefont {X.-Y.}\ \bibnamefont
  {Zhu}},\ }\Doi {10.1016/j.surfrep.2004.09.002} {\bibfield  {journal}
  {\bibinfo  {journal} {Surf. Sci. Rep.},\ }\textbf {\bibinfo {volume} {56}},\
  \bibinfo {pages} {1} (\bibinfo {year} {2004})}\BibitemShut {NoStop}%
\bibitem [{\citenamefont {G\"{u}dde}\ \emph {et~al.}(2006)\citenamefont
  {G\"{u}dde}, \citenamefont {Berthold},\ and\ \citenamefont
  {H\"{o}fer}}]{Gudde2006}%
  \BibitemOpen
  \bibfield  {author} {\bibinfo {author} {\bibfnamefont {J.}~\bibnamefont
  {G\"{u}dde}}, \bibinfo {author} {\bibfnamefont {W.}~\bibnamefont {Berthold}},
  \ and\ \bibinfo {author} {\bibfnamefont {U.}~\bibnamefont {H\"{o}fer}},\
  }\Doi {10.1021/cr050171s} {\bibfield  {journal} {\bibinfo  {journal} {Chem.
  Rev.},\ }\textbf {\bibinfo {volume} {106}},\ \bibinfo {pages} {4261}
  (\bibinfo {year} {2006})}\BibitemShut {NoStop}%
\bibitem [{\citenamefont {Bronner}\ \emph {et~al.}(2011)\citenamefont
  {Bronner}, \citenamefont {Schulze}, \citenamefont {Franke}, \citenamefont
  {Pascual},\ and\ \citenamefont {Tegeder}}]{Bronner2011}%
  \BibitemOpen
  \bibfield  {author} {\bibinfo {author} {\bibfnamefont {C.}~\bibnamefont
  {Bronner}}, \bibinfo {author} {\bibfnamefont {G.}~\bibnamefont {Schulze}},
  \bibinfo {author} {\bibfnamefont {K.~J.}\ \bibnamefont {Franke}}, \bibinfo
  {author} {\bibfnamefont {J.~I.}\ \bibnamefont {Pascual}}, \ and\ \bibinfo
  {author} {\bibfnamefont {P.}~\bibnamefont {Tegeder}},\ }\Doi
  {10.1088/0953-8984/23/48/484005} {\bibfield  {journal} {\bibinfo  {journal}
  {J. Phys.: Condens. Matter},\ }\textbf {\bibinfo {volume} {23}},\ \bibinfo
  {pages} {484005} (\bibinfo {year} {2011})}\BibitemShut {NoStop}%
\bibitem [{\citenamefont {Bronner}\ \emph {et~al.}(2012)\citenamefont
  {Bronner}, \citenamefont {Schulze}, \citenamefont {Hagen},\ and\
  \citenamefont {Tegeder}}]{Bronner2012}%
  \BibitemOpen
  \bibfield  {author} {\bibinfo {author} {\bibfnamefont {C.}~\bibnamefont
  {Bronner}}, \bibinfo {author} {\bibfnamefont {M.}~\bibnamefont {Schulze}},
  \bibinfo {author} {\bibfnamefont {S.}~\bibnamefont {Hagen}}, \ and\ \bibinfo
  {author} {\bibfnamefont {P.}~\bibnamefont {Tegeder}},\ }\Doi
  {10.1088/1367-2630/14/4/043023} {\bibfield  {journal} {\bibinfo  {journal}
  {New J. Phys.},\ }\textbf {\bibinfo {volume} {14}},\ \bibinfo {pages}
  {043023} (\bibinfo {year} {2012})}\BibitemShut {NoStop}%
\bibitem [{\citenamefont {Leyssner}\ \emph {et~al.}(2010)\citenamefont
  {Leyssner}, \citenamefont {Hagen}, \citenamefont {\'{O}v\'{a}ri},
  \citenamefont {Doki\'{c}}, \citenamefont {Saalfrank}, \citenamefont {Peters},
  \citenamefont {Hecht}, \citenamefont {Klamroth},\ and\ \citenamefont
  {Tegeder}}]{Leyssner2010}%
  \BibitemOpen
  \bibfield  {author} {\bibinfo {author} {\bibfnamefont {F.}~\bibnamefont
  {Leyssner}}, \bibinfo {author} {\bibfnamefont {S.}~\bibnamefont {Hagen}},
  \bibinfo {author} {\bibfnamefont {L.}~\bibnamefont {\'{O}v\'{a}ri}}, \bibinfo
  {author} {\bibfnamefont {J.}~\bibnamefont {Doki\'{c}}}, \bibinfo {author}
  {\bibfnamefont {P.}~\bibnamefont {Saalfrank}}, \bibinfo {author}
  {\bibfnamefont {M.~V.}\ \bibnamefont {Peters}}, \bibinfo {author}
  {\bibfnamefont {S.}~\bibnamefont {Hecht}}, \bibinfo {author} {\bibfnamefont
  {T.}~\bibnamefont {Klamroth}}, \ and\ \bibinfo {author} {\bibfnamefont
  {P.}~\bibnamefont {Tegeder}},\ }\Doi {10.1021/jp909684x} {\bibfield
  {journal} {\bibinfo  {journal} {J. Phys. Chem. C},\ }\textbf {\bibinfo
  {volume} {114}},\ \bibinfo {pages} {1231} (\bibinfo {year}
  {2010})}\BibitemShut {NoStop}%
\bibitem [{\citenamefont {Hagen}\ \emph {et~al.}(2010)\citenamefont {Hagen},
  \citenamefont {Luo}, \citenamefont {Haag}, \citenamefont {Wolf},\ and\
  \citenamefont {Tegeder}}]{Hagen2010}%
  \BibitemOpen
  \bibfield  {author} {\bibinfo {author} {\bibfnamefont {S.}~\bibnamefont
  {Hagen}}, \bibinfo {author} {\bibfnamefont {Y.}~\bibnamefont {Luo}}, \bibinfo
  {author} {\bibfnamefont {R.}~\bibnamefont {Haag}}, \bibinfo {author}
  {\bibfnamefont {M.}~\bibnamefont {Wolf}}, \ and\ \bibinfo {author}
  {\bibfnamefont {P.}~\bibnamefont {Tegeder}},\ }\Doi
  {10.1088/1367-2630/12/12/125022} {\bibfield  {journal} {\bibinfo  {journal}
  {New J. Phys.},\ }\textbf {\bibinfo {volume} {12}},\ \bibinfo {pages}
  {125022} (\bibinfo {year} {2010})}\BibitemShut {NoStop}%
\end{thebibliography}
\end{document}